\documentclass[11pt]{article}

\usepackage{ amsmath, amssymb,array}

\newcommand{\Dslash}{\!\not\!\! D}

\newcommand{\dslash}{\!\not\!\partial}

\def\m@th{\mathsurround=0pt }
\def\leftrightarrowfill{$\m@th \mathord\leftarrow \mkern-6mu
	\cleaders\hbox{$\mkern-2mu \mathord- \mkern-2mu$}\hfill
	\mkern-6mu \mathord\rightarrow$}

\def\overleftrightarrow#1{\vbox{\ialign{##\crcr
	\leftrightarrowfill\crcr\noalign{\kern-1pt\nointerlineskip}
	$\hfil\displaystyle{#1}\hfil$\crcr}}}

\begin{document} 

\baselineskip=18pt

%%%%%%%%%%
%%%%%%%%%%    Title page
%%%%%%%%%%

\thispagestyle{empty}
\vspace{20pt}
\font\cmss=cmss10 \font\cmsss=cmss10 at 7pt

\begin{flushright}
% \today \\
% %
UMD-PP-09-038 \\
CERN-PH-TH/2009-085
\end{flushright}

\hfill
\vspace{20pt}

\begin{center}
{\Large \textbf
{Composite Higgs-Mediated FCNC}
}
\end{center}

\vspace{15pt}

\begin{center}
{\large Kaustubh Agashe$\, ^{a}$, Roberto Contino $\, ^{b, c}$} \\
\vspace{15pt}
$^{a}$\textit{Maryland Center for Fundamental Physics,
     Department of Physics,
     University of Maryland,
     College Park, MD 20742, USA}
\\
$^{b}$\textit{Dipartimento di Fisica, Universit\`a di Roma ``La Sapienza'' and \\
INFN Sezione di Roma, I-00185 Roma, Italy}
\\
$^{c}$\textit{CERN, Theory Division, CH 1211, Geneva 23, Switzerland}

\end{center}

\vspace{20pt}

\begin{center}
\textbf{Abstract}
\end{center}
\vspace{5pt} {\small \noindent 
We discuss how, in the presence of higher-dimensional operators,
the Standard Model (SM) fermion masses can be misaligned in flavor space
with the Yukawa couplings to the Higgs boson, even with only one Higgs doublet. 
Such misalignment results in flavor-violating
couplings to the Higgs and hence flavor-changing neutral
current (FCNC) processes from tree-level Higgs exchange. 
We perform a model-independent analysis of such an effect. 
Specializing to the framework of a composite 
Higgs with partially composite SM gauge and fermion fields,
we show that the constraints on the compositeness scale
implied by $\epsilon_K$ can be generically as strong as those from 
the exchange of heavy spin-$1$ resonances
if the Higgs is light and strongly coupled to the new states.
In the special and well motivated case of 
a composite pseudo-Goldstone Higgs, we find that the 
shift symmetry acting on the Higgs forces an alignment
of the fermion mass terms with their Yukawa couplings
at leading order in the fermions' degree of compositeness,
thus implying much milder bounds.
As a consequence of the flavor-violating Higgs couplings,
we estimate $BR(t \to c h)\sim 10^{-4}$ and $BR(h \to tc)\sim 5 \times 10^{-3}$
both for a pseudo-Goldstone (if $t_R$ is fully composite) and for 
a generic composite Higgs.
By virtue of the AdS/CFT correspondence, our results directly apply to 
5-dimensional Randall-Sundrum compactifications.
}

\vfill\eject
\noindent

%%%%%%%%%%
%%%%%%%%%%    Main Text
%%%%%%%%%%

%%%%%%%%%%%%%%%%%%%%%%%%%%%%%%%%%%%%%%%%%%%%%%%%%%%%%%%%%%%%%%%%%%%%%%%%%%%%%%%%%
\section{Introduction}

According to the minimal description of the electroweak symmetry breaking (EWSB) 
in the Standard Model (SM), the Yukawa couplings of the 
up and down quarks to the Higgs boson
are exactly aligned in flavor space with their mass terms, 
so that no flavor violation can arise mediated by the Higgs.
There are however serious reasons to think of the Standard Model as an effective
theory  with a $\sim\,$TeV cutoff, and this simple picture 
could be dramatically modified by the New Physics.
The leading flavor-violating 
contribution in the Higgs couplings to fermions can be
parametrized by dimension-6 operators in the effective Lagrangian 
with more powers of the Higgs doublet than at the renormalizable level.

At first sight the Higgs contribution to  $\Delta F=2$ neutral currents is subdominant
compared to that originating from generic dimension-6 four-fermion operators, 
as for example those arising from the exchange of new heavy vectors,
since it requires two flavor-violating Higgs vertices, and as such it naively
corresponds to a dimension-8 effect.
However, as we will show in this paper, if the Higgs boson is light and strongly coupled 
to the new dynamics then its contribution can 
be comparable to dimension-6 effects
and imply strong bounds on the scale of the new states.
In this limit it will also parametrically dominate over $\Delta F=2$ contributions
generated through the exchange of the $Z$ boson, which is light yet
weakly coupled to the new physics.
The Higgs contribution to $\Delta F=1$ transitions will instead be largely negligible
compared to the $Z$ exchange, as further suppressed by a Yukawa coupling factor
at the flavor-preserving vertex.

A strongly-coupled light Higgs can naturally arise as the 
composite pseudo-Goldstone boson (PGB) of the new dynamics 
responsible for the electroweak symmetry breaking~\cite{GK}.
This possibility has recently attracted much attention
as it resolves the Planck-weak hierarchy problem of the Standard
Model while still being compatible with the precision tests
performed at LEP (see, for example,~\cite{Giudice:2007fh}). 
In addition, if the SM quarks couple linearly to the new dynamics
through composite fermionic operators~\cite{Kaplan:1991dc}, then the hierarchies in their masses
and mixing angles can also be elegantly explained as the effect of a
renormalization group evolution using only mild differences in the scaling dimensions 
of the operators.
Remarkably, this scenario is explicitly realized in the 4D duals of 5D warped 
compactifications~\cite{Randall:1999ee, Gherghetta:2000qt, Contino:2003ve, Agashe:2004rs} 
as per the AdS/CFT correspondence~\cite{Maldacena:1997re, Arkani-Hamed:2000ds}.
Numerous theoretical studies on flavor-violating processes in such 5D realizations
of composite Higgs models (and their 4D deconstructions) have been performed,
mainly focussing on $\Delta F=2$ transitions mediated by the tree-level exchange
of heavy vectors or on $\Delta F = 1$ rare decays arising at tree- and
one-loop level: see ~\cite{Huber:2003tu, Agashe:2004cp, Csaki:2008zd, %
Casagrande:2008hr, Albrecht:2009xr} and 
references therein.

In this paper we show that the tree-level exchange of the Higgs boson
can lead to quite strong constraints in $\Delta F=2$ processes.
In the following sections we first present a general model-independent
operator analysis, then we specialize to the case of a composite Higgs 
(assuming linear couplings of the SM quarks to
the strong dynamics), with a dedicated analysis of the
pseudo-Goldstone limit. 

Flavor-violating Higgs couplings were independently investigated in 
refs.~\cite{Casagrande:2008hr} and \cite{Blanke:2008zb,Albrecht:2009xr}
in the context of specific 5D Randall-Sundrum models with a (non-PGB) Higgs doublet 
localized on or at the vicinity of the infrared brane. 
In both constructions, as a consequence of the Higgs localization, 
the effects were found to be small.
In either case a full, general analysis 
of Higgs-mediated $\Delta F =2$ neutral currents
was not carried through.
Finally, similar flavor-violating effects from the Higgs sector have been
considered in~\cite{Babu:1999me,Giudice:2008uua}, although in a different context.

\section{Model-independent analysis of
higher-dimensional operators}

In this section we present a model-independent analysis of the flavor-violating effects
induced by the tree-level Higgs exchange, and derive the corresponding experimental 
constraints.~\footnote{Similar operator analyses
can be found in~\cite{Buchmuller:1985jz,delAguila:2000aa,AguilarSaavedra:2009mx},
though no experimental bound was there derived.}
We focus on the down-type quark sector since it gives the 
strongest constraints, and leave to the reader the straightforward generalization to other sectors. 
At the level of dimension 6, there are four
operators in the effective Lagrangian for a light Higgs doublet $H$ which can lead
to flavor-violating couplings to the SM down-type quarks. The first 
is a non-derivative operator
\begin{equation}
O_y = \frac{\tilde{y}^d_{ i j }}{\Lambda^2}\;  \bar{ q}_{L\, i}  H d_{ R \, j } (H^{ \dagger } H ) + h.c. 
\label{eq:Oy}
\end{equation}
while the other three are Higgs-dependent modifications of the quark kinetic terms:
\begin{equation}
\label{eq:otherO}
\begin{split}
O_{q} & = \frac{\kappa^{q}_{ij}}{\Lambda^2} \;  \bar{ q}_{ L \, i \,} i\Dslash q_{L\, j} (H^\dagger H) + h.c. \\[0.2cm]
O_q^\prime & = \frac{\kappa^{\prime\, q}_{ij}}{\Lambda^2} \;  \bar{ q}_{ L \, i\, } H H^\dagger i \Dslash q_{L\, j} + h.c.\\[0.2cm]
O_{d} & = \frac{\kappa^d_{ij}}{\Lambda^2} \;   \bar{ d}_{R\, i\,} i\Dslash d_{R\, j}  (H^\dagger H) + h.c. \, . 
\end{split}
\end{equation}
Here $i, j$ denote generation indices, $\tilde y_{ij}$ and $\kappa_{ij}$ are generic complex coefficients,
and $\Lambda$ stands for the mass scale of the New Physics.
Notice that the additional independent operators
\begin{equation}
\label{eq:Zoperators}
 \bar{ q}_{L\, i} \gamma^\mu q_{ L \, j } (H^\dagger i {\overleftrightarrow{D_\mu}} H ) + h.c. \qquad 
 \bar{ q}_{L\, i} \gamma^\mu T^a q_{L\, j} (H^\dagger i{\overleftrightarrow{D_\mu}} T^a H) + h.c. \qquad
 \bar{ d}_{R\, i} \gamma^\mu d_{ R \, j } (H^\dagger i {\overleftrightarrow{D_\mu}} H ) + h.c.
\end{equation}
with $H^\dagger\overleftrightarrow{D_\mu} H \equiv H^\dagger D_\mu H - (D_\mu H)^\dagger H$, 
do not modify the couplings of the Higgs boson (though they do  modify the couplings
of the $Z$), and are thus not relevant here.

As noticed in~\cite{AguilarSaavedra:2009mx}, the operators of eq.(\ref{eq:otherO}) can be 
rewritten in terms of $O_y$ by means of a field redefinition,~\footnote{Such field redefinition 
is equivalent to using the
classical equations of motion upon the higher-dimensional operators~\cite{OSEFT}. 
We thank J.A.~Aguilar-Saavedra for a clarifying discussion on this point.} so that
\begin{equation}
\label{eq:redef}
\tilde y_{ij}^d \to  \tilde y_{ij}^d + \big( \kappa^q \cdot y^d \big)_{ij} +  
\big( \kappa^{\prime\, q} \cdot y^d \big)_{ij} + \big(  y^d \cdot  \kappa^{d\, \dagger} \big)_{ij}\, ,
\end{equation}
where $y^d$ is the quark down Yukawa matrix.  It is however convenient 
to distinguish between $O_y$ and the derivative operators in eq.(\ref{eq:otherO}), as
different spurionic transformation rules can be assigned to their coefficients
\begin{equation}
\begin{gathered}
y^d \to V_L\, y^d\, V_R^\dagger \, , \qquad  \tilde y^d \to V_L\, \tilde y^d\, V_R^\dagger \\[0.3cm]
\kappa^q \to V_L\, \kappa^q\, V_L^\dagger \, , \qquad \kappa^{\prime q} \to V_L\, \kappa^{\prime q}\, V_L^\dagger \, ,
\qquad \kappa^{d} \to V_R\, \kappa^{d}\, V_R^\dagger
\end{gathered}
\end{equation}
under flavor $SU(3)_{L,R}$ rotations $q_{L\, i} \to (V_{L})_{ij}\, q_{L\, j}$, $d_{R\, i} \to (V_{R})_{ij}\, d_{R\, j}$.
In particular, we will see in section~\ref{sec:PGB} that in theories where the Higgs is a pseudo-Goldstone boson,
$\tilde y^d$ can be naturally aligned with $y^d$ (so that the only flavor spurion transforming as
a $(3,3)$ under $SU(3)_L\times SU(3)_R$  is the Yukawa coupling). In this case
flavor-violating effects arise only from the derivative operators.

By use of eq.(\ref{eq:redef}), after expanding around the EWSB vacuum ($H^0 = v + h(x)$, with $v = 174\,\text{GeV}$), 
and keeping at most terms linear in the Higgs field $h(x)$, the effective Lagrangian involving
down-type quarks reads:
\begin{equation}
\begin{split}
{\cal L} = & \bar{d}_{L} i\!\dslash d_{L} + \bar{d}_{R} i\!\dslash d_{R} -
  v\, \bar{d}_{ L \, i }  d_{ R \, j }  
  \left[ y_{ i j }^d - \big( \tilde{y}^d + \kappa^q y^d + \kappa^{\prime\, q} y^d + y^d \kappa^{d\,\dagger} \big)_{ij} 
   \frac{v^2}{\Lambda^2}  \right] \\
  & - h\, \bar{d}_{ L \, i }  d_{ R \, j }  
  \left[ y_{ i j }^d - 3\, \big( \tilde{y}^d + \kappa^q y^d + \kappa^{\prime\, q} y^d + y^d \kappa^{d\,\dagger} \big)_{ij}  
   \frac{v^2}{\Lambda^2} \right] + h.c. \, .
\end{split}
\end{equation}
The down quark mass matrix
\begin{equation}
m^d_{ i j }  =  y^d_{ i j } v - \big( \tilde{y}^d + \kappa^q y^d + \kappa^{\prime\, q} y^d + y^d \kappa^{d\,\dagger} \big)_{ij} 
  \frac{ v^3 }{ \Lambda^2 }\, ,
\end{equation}
can be diagonalized as usual through a bi-unitary transformation, 
$q_{ L \, i } \to \left( D_L \right)_ { i j } q_{ L \, j }$ and $d_{ R \, i } \to \left( D_R \right)_ { i j } d_{ R \, j }$,
so that $m^{d}_i \delta_{ij} = (D_L^\dagger m^d D_R)_{ij}$.
In this mass-eigenstate basis the couplings to the physical Higgs boson $h$ are not flavor diagonal:
\begin{equation}
{\cal L} = \bar{d}_{L} i\!\dslash d_{L} + \bar{d}_{R} i\!\dslash d_{R} -
 \bar{d}_{ L \, i } m^d_{i}  d_{ R \, i } - h \, \bar{d}_{ L \, i }  d_{ R \, j }  \left( \frac{ m^d_{i}}{v}\delta_{ij} 
 - \hat{y}^d_{ i j }\,  \frac{v^2}{\Lambda^2} \right) + h.c.
\end{equation}
where we have defined:
\begin{equation}
\label{eq:yhatdef}
\hat y_{ij}^d \equiv  \, - 2 \left[ D^\dagger_L \cdot \tilde y^d \cdot D_R \right]_{ij}
 - 2 \left[ D^\dagger_L \cdot \left( \kappa^{q} + \kappa^{\prime\, q}  \right) \cdot D_L \right]_{ij}  \frac{m^d_{j}}{v} 
  - 2 \,  \frac{m^d_{i}}{v} \left[ D^\dagger_R \cdot \kappa^{d\,\dagger} \cdot D_R
 \right]_{ij} 
\end{equation}
and neglected higher-order terms in $(v/\Lambda)$.
As a consequence, the tree-level exchange of the Higgs boson will generate $\Delta F =2$ transitions
at low energy. 

To illustrate the importance of the effect we consider for example
the contribution to $K\bar{K}$ mixing: by integrating out the Higgs boson one obtains
the low-energy Lagrangian ($\alpha$ denotes a color index)
\begin{gather}
{\cal L}_{ \Delta S = 2 } = {\cal O}_2 C_2 + \tilde{ {\cal O} }_2 \tilde{C}_2 + {\cal O}_4 C_4 + h.c. \\[0.5cm]
{\cal O}_4  \equiv \left( \overline{ s^{\alpha}_L } d_{ R \, \alpha } \right)   
  \left( \overline{ s^{ \alpha }_R } d_{ L \, \alpha } \right) , \qquad
{\cal O}_2 \equiv \left( \overline{ s^{ \alpha }_R } d_{ L \,\alpha } \right)^2   ,\qquad
\tilde{ {\cal O} }_2 \equiv \left( \overline{ s^{ \alpha }_L } d_{ R \, \alpha } \right)^2 \, ,
\end{gather}
with
\begin{equation}
\left( C_4, \; C_2, \; \tilde{C}_2 \right)  =  
\frac{1}{m_h^2} \left( \frac{ v^2 }{ \Lambda^2 } \right)^2 
\left( \hat{y}^d_{ 12 } \hat{y}^{ d \; \ast }_{ 21 }, 
\; \frac{1}{2}\, (\hat{y}^{d\, *}_{ 12 })^2 , \;  \frac{1}{2}\, (\hat{y}^{d}_{ 2 1 })^2 \right)\, .
\end{equation}
Since these $4$-fermion operators are generated through the exchange of the Higgs, we must then apply the
experimental constraint on the Wilson coefficients $C_i$ renormalized at the Higgs mass scale. 
Using the RGE equations from reference \cite{Bagger:1997gg} to evolve the experimental constraints
reported by the UTFit collaboration \cite{Bona:2007vi}, and choosing a reference Higgs mass
$m_h = 200\,$GeV,~\footnote{The bounds are only logarithmically sensitive to the renormalization
scale.} we find
\begin{equation}
\text{Im} \left( C_4, \; C_2, \; \tilde{C}_2
\right) \, (\mu = m_h = 200\,\text{GeV})  \lesssim 
\frac{1}{ \Big\{ \left( 1.4, \; 0.72, \; 0.72 \right)
\times 10^5 \; \hbox{TeV} \Big\} ^2 }\, .
\end{equation}
By turning on one operator at a time (which gives a rough account of the global constraint if these contributions 
are uncorrelated), the above bound implies
\begin{equation}
\Lambda \gtrsim  (145, 88, 88) \, \text{TeV}\, \sqrt{ \frac{ 200 \, \text{GeV} }{ m_h } }
\cdot \left[ \text{Im} \left( \hat{y}^d_{ 12 } \hat{y}^{ d \; \ast }_{ 21 }, \;
(\hat{y}^{ d\, *}_{ 12 })^2, \; (\hat{y}^{d}_{ 2 1 })^2 \right) \right]^{1/4}\, .
\label{model-indep}
\end{equation}

Following the same steps as above, it is straightforward to derive the analogous
bounds on  $\Lambda$ from the $B_d\bar B_d$ and $B_s\bar B_s$ systems. 
We leave all the intermediate formulas to the appendix  and quote here only
the final results: we find
\begin{equation}
\Lambda \gtrsim  (15, 10, 10) \, \text{TeV}\, \sqrt{ \frac{ 200 \, \text{GeV} }{ m_h } }
\cdot \left[ \text{Im} \left( \hat{y}^d_{ 13 } \hat{y}^{ d \; \ast }_{ 31 }, \;
(\hat{y}^{ d\, *}_{ 13 })^2, \; (\hat{y}^{d}_{ 3 1 })^2 \right) \right]^{1/4}
\label{model-indepdB}
\end{equation}
from $B_d\bar B_d$ mixing, and
\begin{equation}
\Lambda \gtrsim  (5.2, 3.4, 3.4) \, \text{TeV}\, \sqrt{ \frac{ 200 \, \text{GeV} }{ m_h } }
\cdot \left[ \text{Im} \left( \hat{y}^d_{ 23 } \hat{y}^{ d \; \ast }_{ 32 }, \;
(\hat{y}^{ d\, *}_{ 23 })^2, \; (\hat{y}^{d}_{ 3 2 })^2 \right) \right]^{1/4}\, 
\label{model-indepdBdS}
\end{equation}
from $B_s\bar B_s$ mixing.

\section{Flavor violation from a composite Higgs}

The constraint on the New Physics scale $\Lambda$ derived above can be made more explicit
by indicating the origin of the higher-dimensional operators in eqs.(\ref{eq:Oy}),(\ref{eq:otherO}).
Here we consider the motivated scenario in which the Higgs doublet arises 
as a bound state of a new strongly-interacting dynamics.
We assume that the SM fermions are linearly coupled to the strong sector
through composite operators $O_{L,R}$~\cite{Kaplan:1991dc}
\begin{equation}
\label{lincouplings}
\lambda_L\, \bar\psi_L O_R + \lambda_R\, \bar\psi_R O_L + h.c. \, , 
\end{equation}
like in the 4D duals of 5D warped compactifications.
The SM Yukawa term and the higher-order operators of eqs.(\ref{eq:Oy}),(\ref{eq:otherO}) 
then arise at low energy by expanding the two-point Green functions of $O_{L,R}$ in powers of the 
Higgs field.

The coefficients $\tilde y^d$, $\kappa$ and the SM Yukawa coupling $y^d$ can
be estimated by means of Naive Dimensional Analysis (NDA) as follows:
\begin{equation}
\label{eq:NDAestimates}
y^d \sim y_* \, \frac{\lambda_L \lambda_R}{16\pi^2} 
\qquad\quad
\tilde y^d \sim y_*^3 \, \frac{\lambda_L \lambda_R}{16\pi^2}
\qquad\quad
\kappa^{q},\kappa^{\prime\, q} \sim  y_*^2 \, \frac{\lambda_L^2}{16\pi^2} 
\qquad\quad
\kappa^{d} \sim  y_*^2 \, \frac{\lambda_R^2}{16\pi^2}\, ,
\end{equation}
where the coupling of the composite Higgs to the other strong states, $y_*$,
can be as large as $4\pi$, this case
corresponding to a maximally  strongly coupled dynamics. 

We further assume that the hierarchy in the SM Yukawa couplings $y^d_{ij}$ entirely originates from
the hierarchy in the couplings $\lambda_{L,R}$, as for example the effect of their RG evolution,
whereas the strong sector is substantially flavor anarchic.
This has been dubbed in the literature as the ``anarchic scenario'',
and has been studied extensively in the 5D warped
framework, see~\cite{Huber:2003tu, Agashe:2004cp,Csaki:2008zd, Casagrande:2008hr, Albrecht:2009xr} 
and references therein.
Then, making the flavor structure explicit, the matrix $\tilde y^d$ will have the same
hierarchical structure of $y^d$ in flavor space, but it will not be exactly aligned with it in general:
\begin{equation}
\tilde y^d_{ij} = y_*^2\; a_{ij} \times y^d_{ij} \qquad \text{(no sum over $i$, $j$)}\, ,
\end{equation}
where $a_{ i j }$ is an anarchic matrix with $O(1)$ entries. By applying the above estimates to 
eq.(\ref{eq:yhatdef}) one finds, in the mass eigenstate basis 
\begin{equation}
\label{eq:yhatest}
\begin{split}
\hat y_{ij}^d \sim \,  \frac{2\, y_*^2}{16\pi^2}\,\bigg[
  &  y_*\, (D_L^\dagger)_{il}\, (\lambda_L)_l \, (\lambda_R)_m \, (D_R)_{mj} 
    + (D_L^\dagger)_{il}\, (\lambda_L)_l \, (\lambda_L)_m \, (D_L)_{mj} \, \frac{m^d_j}{v} \\
  & + \frac{m^d_i}{v}\, (D_R^\dagger)_{il} \, (\lambda_R)_l \, (\lambda_R)_m \, (D_R)_{mj}  \bigg]  \, ,
\end{split}
\end{equation}
where we have omitted $O(1)$ factors.
As it will be explicitly shown in the next section, the second and third terms,
which have their origin in the derivative operators, are always subdominant or at best
of the same order of the first term,  which arises from the operator $O_y$.
\footnote{The occurrence of non-universal shifts in the Higgs couplings as
a consequence of corrections to the SM fermion kinetic terms was already noticed 
in ref.~\cite{Giudice:2007fh} in the general context of a composite Higgs.
Formulas for the modified Higgs couplings analogous to the last two terms
and the first term of eq.(\ref{eq:yhatest}) were reported, respectively, in 
refs.\cite{Casagrande:2008hr,Blanke:2008zb} and \cite{Blanke:2008zb}
for the case of specific 5D Randall-Sundrum models with the Higgs localized on or at the
vicinity of the infrared brane.
See reference~\cite{Azatov} for a further discussion on the  
effect of the first term in both bulk and brane Higgs 5D scenarios.}

We thus concentrate on the dominant effect from $O_y$ and drop the  
last two terms of eq.(\ref{eq:yhatest}).
As a final simplification, we assume that the left
rotation matrix has entries of the same order as those of the Cabibbo-Kobayashi-Maskawa matrix:
\begin{equation}
\label{eq:LHest}
(D_L)_{ij} \sim (V_{CKM})_{ij} \, .
\end{equation}
This in turn, combined with the anarchy assumption and the estimate of the Yukawa matrix
in eq.(\ref{eq:NDAestimates}), fixes the form of the right couplings $\lambda_R$ and rotation matrix $D_R$:
\begin{equation}
\label{eq:RHest}
(\lambda_R)_i \sim \frac{m_i}{y_* v\, (\lambda_L)_i}  \, ,
\quad\qquad 
(D_R)_{ij} \sim \left(\frac{m_i}{m_j}\right) \frac{1}{(D_L)_{ij}} \quad \text{for $i<j$} \, .
\end{equation}
Using the above estimates and specializing to $K\bar K$ mixing,
we find that ($\lambda_C = 0.22$ denotes the Cabibbo angle)
\begin{equation}
\hat y_{12}^d \sim 2\, y_*^2\, \frac{m_s}{v} \, \lambda_C \, ,
\label{eq:yhat12}
\end{equation}
which arises as the result of 
several equally important terms in the sum over $l,m$ of eq.(\ref{eq:yhatest}), with flavor violation arising
either from the vertex $\tilde y^d$, or from the rotation to the mass eigenstates basis, or from both.
A similar estimate can be derived for $\hat y_{21}^d$. 
Using $m_{ d, s }= 3, 65$ MeV, i.e. the value of the quark masses renormalized at the Higgs mass 
scale $m_h=200$ GeV, and assuming $O(1)$ CP-violating phases, one has
\begin{equation}
\label{eq:finalbound}
\Lambda \gtrsim  (1.9, 1.1, 1.1) \, \text{TeV}\times y_* \times
 \sqrt{ \frac{ 200 \, \text{GeV} }{ m_h } }\, .
\end{equation}
Considering that $\Lambda$ should be identified with the mass scale of the fermionic
resonances of the strong sector, and that $y_*$ can in principle be as large as $4\pi$, 
the above constraint is rather strong.

It is interesting to compare with the constraints on $\Delta F=2$ FCNCs that arise from the 
tree-level exchange of heavy colored vectors (such as KK or composite gluons) and from the $Z$ boson.
In the latter case the leading New Physics effects are encoded by the dimension-6 operators
in eq.(\ref{eq:Zoperators}).
After matching to the four-fermion low-energy effective Lagrangian, a rough NDA estimate
of the size of a generic Wilson coefficient (evaluated at the matching scale and
before rotating to the mass eigenstate basis) gives
\begin{equation}
\begin{aligned}
C_i(m_h) & \sim \frac{y_*^2}{m_h^2} \left( \frac{\lambda^2}{16\pi^2} \right)^2 
\left( \frac{y_*^2 v^2}{\Lambda^2} \right)^2 & \qquad &\text{from Higgs exchange} \\[0.2cm]
C_i(M_Z) & \sim \frac{g^2}{M_Z^2} \left( \frac{\lambda^2}{16\pi^2} \right)^2 
\left( \frac{g_*^2 v^2}{\Lambda^2} \right)^2 & &\text{from $Z$ exchange} \\[0.2cm]
C_i(M_*) & \sim \frac{g_*^2}{\Lambda^2} \left( \frac{\lambda^2}{16\pi^2} \right)^2 
 & &\text{from heavy vector exchange}\, .
\end{aligned}
\end{equation}
Here $g_*$ ($\sim y_*$) denotes the typical coupling of the heavy resonances of the strong sector, 
while $\lambda$ stands for $\lambda_L$ or $\lambda_R$.
The first estimate, in particular, agrees with the more refined one in eq.(\ref{eq:yhatest}).
This shows that the $Z$ exchange is always suppressed by
a factor $(g_*^2 v^2/\Lambda^2)$ compared to the heavy vector exchange, 
and this is the reason why it has been usually neglected in the literature.
The Higgs exchange, on the other hand,  while suffering from the same suppression,
has a further enhancement factor $(y_*^2 v^2/m_h^2)$ which 
is large if the Higgs is light and strongly coupled to the new states. 
In this way  it can become comparable to the genuine dimension-6 effects.

In the case of $\Delta S=2$ transitions the dominant contribution from heavy vectors 
to $\epsilon_K$ is from $C_4$, and it leads to the following bound on the heavy vector mass  
(assuming an $O(1)$ CP-violating phase)~\cite{Agashe:2004cp,Csaki:2008zd,Agashe:2008uz}:
\begin{equation}
\label{eq:vectorbound}
\Lambda \gtrsim 10 \,  \text{TeV} \times \left( \frac{g_*}{y_*} \right)^2 .
\end{equation}
This is comparable with the bound on the New Physics scale of eq.(\ref{eq:finalbound}) from 
the Higgs exchange for $y_* \sim g_* \sim 5$ and $m_h = 200\,\text{GeV}$.
\footnote{Notice however, that in principle
the two New Physics scales entering eq.(\ref{eq:finalbound}) and eq.(\ref{eq:vectorbound}) 
might be different,  as they naively correspond to the mass of respectively the heavy 
vectorial and fermionic resonances.}
Furthermore, the above constraint becomes weaker by making 
the coupling of the heavy vector 
$g_*$ smaller,  or $y_*$ larger.~\footnote{For fixed SM Yukawa couplings this latter 
limit is equivalent to making $(\lambda_L\lambda_R)$ smaller.}
The bound of eq.(\ref{eq:finalbound}), instead, becomes stronger for larger $y_*$, and 
in this sense the two are complementary.

The constraints that follow from $\Delta B=2$ processes
due to the Higgs exchange are less severe than those obtained above from CP violation in
$K\bar K$ mixing.
Under the same assumptions which led to eq.~(\ref{eq:yhat12})
and using $m_b = 3\,$GeV (renormalized at the Higgs mass 
scale $m_h=200$ GeV), one finds
\begin{align}
& \hat y^d_{ 1 3 } \sim 2\, y_*^2\, \frac{m_b}{v} \lambda_C^3  \, , &  
\hat y^d_{ 31 } \sim 2\, y_*^2\, \frac{m_b}{v} \left( \frac{m_d}{m_b \lambda_C^3} \right)\, , \\[0.2cm]
%%%
&\hat y^d_{ 2 3 } \sim 2\, y_*^2\,  \frac{m_b}{v}\, \lambda_C^2\, , & 
\hat y^d_{ 32 } \sim 2\, y_*^2\, \frac{m_b}{v} \left( \frac{m_s}{m_b \lambda_C^2} \right)\, .
\end{align}
After substituting these expressions in eqs.(\ref{model-indepdB}) and (\ref{model-indepdBdS}) 
and assuming $O(1)$ phases, one obtains the following constraints:
\begin{equation}
\label{BBbound}
\Lambda \gtrsim  (480, 190, 570) \, \text{GeV}\times y_* \times
 \sqrt{ \frac{ 200 \, \text{GeV} }{ m_h } }
\end{equation}
and
\begin{equation}
\label{BsBsbound}
\Lambda \gtrsim  (360, 135, 420) \, \text{GeV}\times y_* \times
 \sqrt{ \frac{ 200 \, \text{GeV} }{ m_h } }\, ,
\end{equation}
respectively from $B_d \bar B_d$ and $B_s \bar B_s$ mixings.

\section{PGB Composite Higgs and flavor alignment}
\label{sec:PGB}

There is an important class of composite Higgs models where the strong constraint
on $\Lambda$ derived in the previous section does not hold in general: these are theories in which
the Higgs, rather than being an ordinary bound state, is a pseudo-Goldstone boson 
associated to the spontaneous breaking of a global symmetry $G$ of the strong sector~\cite{GK}.
In that case the shift symmetry acting on the Higgs dictates the structure of the 
higher-order Higgs-dependent operators. 

The simplest possibility is that the operators $O_{L}$, $O_R$ in eq.(\ref{lincouplings})
have definite quantum numbers under the global symmetry $G$,  and transform as  
representations $r_{L}$, $r_R$ (for all three generations).
It is possible then to ``uplift'' the SM fermions to complete representations $r_L$, $r_R$ of $G$,
$q_L \to \psi_L$, $d_R \to \psi_R$, the extra components being non-physical spurionic fields.
In this way the Higgs dependence of all non-derivative operators must resum to a polynomial $P$
of the sigma model field $\Sigma = e^{i h/f}$ (where $f$ is the analog of the pion decay constant): 
\begin{equation}
\bar q_L^i  H \left( y^d_{ij} + \tilde y^d_{ij} \frac{H^\dagger H}{\Lambda^2} + \cdots \right) d_R^j
\longrightarrow \bar \psi_L^i  P_{ij}(\Sigma)  \psi_R^j
\end{equation}
where $i,j$ are flavor indices. The polynomial $P$  transforms as a $r_L \times r_R$.
If its projection over the physical fields $q_L$, $d_R$ contains only one term (transforming as 
${2}_{1/2}$ under SU(2)$_L \times$U(1)$_Y$), 
then the flavor dependence in the right-hand-side of the
above equation factorizes and all higher-order terms
in the Higgs field expansion are aligned: 
$\bar\psi_L^i  P_{ij}(\Sigma)  d_R^j = y^d_{ij}\, \bar\psi_L^i  P(\Sigma)  \psi_R^j$.
In particular, $\tilde y_{ij}^d$ is aligned with $y_{ij}^d$ and the constraints of
eqs.(\ref{eq:finalbound}),(\ref{BBbound}),(\ref{BsBsbound}) do not hold.
On the other hand, if the projection of $P$ over $q_L$, $d_R$ contains more than one term,
as for example if any of the SM fermion is coupled (with
similar strength) to more than one composite operator,
than the alignment in flavor space is broken and the same bounds as for a non-PGB Higgs apply.

An explicit example will best illustrate this general result:~\footnote{See 
for example ref.~\cite{Contino:2006qr}
for an explicit realization in the context of a 5D warped model.}
Consider a strong sector with 
$G =$ SO(5)$\times$U(1)$_X$ spontaneously broken to SO(4)$\times$U(1)$_X$,
where SO(4)$\sim$SU(2)$_L\times$SU(2)$_R$ and $Y = T^3_R+X$.
Massless excitations around the SO(4) vacuum are parametrized by the 
Goldstone field $\Sigma = e^{i h/f}$, which transforms as a $5$ of SO(5).
Both operators $O_{L}$, $O_R$ are $5_{-1/3}$ of SO(5)$\times$U(1)$_X$, 
where $5 = (1,1) + (2,2)$ under SU(2)$_L\times$SU(2)$_R$.
Accordingly, each $q_L$ and $d_R$ can be uplifted to a full $5_{-1/3}$ of SO(5)$\times$U(1)$_X$,
respectively denoted as $\psi_L$ and $\psi_R$, so that $d_R$ is the $(1,1)$ inside $\psi_R$
and $q_L$ is the  $T^3_R = -1/2$ component of the $(2,2)$ inside $\psi_L$.
Then, the polynomial $P(\Sigma)$ transforms as a $5\times 5$ and its projection over
the physical fields $d_R$, $q_L$ contains only one term, hence only one flavor structure:
\begin{equation}
\bar \psi_L^i  P_{ij}(\Sigma) \psi_R^j = y^d_{ij}\, \bar \psi_L^i  \Sigma \Sigma^T  \psi_R^j
 = y^d_{ij}\, \sin (h/f) \cos (h/f)\, \bar q_L^i \hat H d_R^j \, ,
\end{equation}
where we have defined
\begin{equation}
\hat H = \frac{1}{h} \begin{bmatrix} h^1 + i h^2 \\ h^3 + i h^4 \end{bmatrix} \, ,
\qquad h = \sqrt{(h^a)^2}\, .
\end{equation}
This shows how the shift symmetry acting on the Higgs forces all 
the non-derivative operators with higher powers of the Higgs field to be aligned
in flavor space with the SM Yukawa term. 
On the other hand, there is still a flavor-universal shift in the couplings of the Higgs boson
of order $v^2/f^2$.

In spite of the flavor alignment in the non-derivative operators,
flavor violation in the Higgs couplings will still occur 
due to the derivative operators of eq.(\ref{eq:otherO}).
Starting from eq.(\ref{eq:yhatest}) and concentrating on the last two terms, it is straightforward to
derive the estimate for $\hat y_{12}^d$ and $\hat y_{21}^d$ relevant for $K\bar K$ mixing
assuming eqs.(\ref{eq:LHest}), (\ref{eq:RHest}). We find:
\begin{equation}
\begin{split}
\hat y_{12}^d \sim & 2 y_*^2 \left( \frac{m_s}{v}\, \frac{(\lambda_L)_1(\lambda_L)_2}{16\pi^2}  
 + \frac{m_d}{v}\, \frac{(\lambda_R)_1(\lambda_R)_2}{16\pi^2} \right)\\[0.4cm]
\hat y_{21}^d \sim & 2 y_*^2 \left( \frac{m_d}{v}\, \frac{(\lambda_L)_1(\lambda_L)_2}{16\pi^2}
 + \frac{m_s}{v}\, \frac{(\lambda_R)_1(\lambda_R)_2}{16\pi^2} \right)\, .
\end{split}
\end{equation} 
The first term in each of the above formulas is maximized in the limit of $b_L$ fully composite
(i.e. for $(\lambda_L)_3 \to 4\pi$), while the second term is maximized for $b_R$ fully composite
($(\lambda_R)_3 \to 4\pi$).
For $b_L$ fully composite the strongest constraint on $\Lambda$ comes from $C_2\propto (\hat y_{12}^d)^2$:
\begin{equation}
\hat y_{12}^d \sim  2\, y_*^2\, \frac{m_s}{v}\, \lambda_C \left[\frac{(\lambda_L)_2}{4\pi}\right]^2 
  \sim  2\, y_*^2\, \frac{m_s}{v}\, \lambda_C^5
\quad \Longrightarrow \quad
\Lambda \gtrsim 55 \, \text{GeV}\times y_* \times \sqrt{\frac{200\,\text{GeV}}{m_h}}\, .
\end{equation}
Compared to the estimate of $\hat y_{12}^d$ from the non-derivative operator in eq.(\ref{eq:yhat12}), 
it is clear that the effect of the derivative operators is at best suppressed by a factor
$\zeta_{s_L}^2 = ((\lambda_L)_2/4\pi)^2$, where $\zeta_{s_L}$ corresponds to the degree of compositeness of $s_L$.
Similarly, for $b_R$ fully composite the strongest constraint on $\Lambda$ comes from 
$\tilde C_2\propto (\hat y_{21}^d)^2$:
\begin{equation}
\hat y_{21}^d \sim  2\, y_*^2\, \frac{m_s}{v}\, \lambda_C \left[\frac{(\lambda_R)_2}{4\pi}\right]^2 
 \sim 2\, y_*^2\, \frac{m_s}{v}\, \lambda_C \left( \frac{m_s}{m_b \lambda_C^2} \right)^2
\quad \Longrightarrow \quad
\Lambda \gtrsim 510\, \text{GeV}\times y_* \times \sqrt{\frac{200\,\text{GeV}}{m_h}}\, .
\end{equation}
Again, compared to eq.(\ref{eq:yhat12})
the effect of the derivative operators is at best suppressed by a factor
$\zeta_{s_R}^2 = ((\lambda_R)_2/4\pi)^2$, with $\zeta_{s_R}$ equal to the degree of compositeness of $s_R$.

The suppression from the degree of compositeness of the SM quarks is instead 
absent in 
$\Delta B=2$ processes in the limit of either $b_L$ or $b_R$ being fully 
composite, in which case the constraint from derivative operators becomes 
as important as that from $O_y$.
In the case of $B_d\bar B_d$ mixing one gets
\begin{equation}
\begin{split}
\hat y_{13}^d \sim & 2 y_*^2 \left( \frac{m_b}{v}\, \frac{(\lambda_L)_1(\lambda_L)_3}{16\pi^2}  
 + \frac{m_d}{v}\, \frac{(\lambda_R)_1(\lambda_R)_3}{16\pi^2} \right)\\[0.4cm]
\hat y_{31}^d \sim & 2 y_*^2 \left( \frac{m_d}{v}\, \frac{(\lambda_L)_1(\lambda_L)_3}{16\pi^2}
 + \frac{m_b}{v}\, \frac{(\lambda_R)_1(\lambda_R)_3}{16\pi^2} \right)\, ,
\end{split}
\end{equation} 
so that the strongest constraint for $b_L$ and $b_R$ fully composite respectively comes
from $C_2\propto (\hat y_{13}^d)^2$ and $\tilde C_2\propto (\hat y_{31}^d)^2$:
\begin{align}
\hspace*{2cm} 
\hat y_{13}^d & \sim  2\, y_*^2\, \frac{m_b}{v}\, \lambda_C^3 
 & \Longrightarrow  \quad
 \Lambda  \gtrsim 190\, \text{GeV}\times y_* \times \sqrt{\frac{200\,\text{GeV}}{m_h}}
\phantom{\, .}
\hspace*{2cm} 
\\[0.3cm]
 \hat y_{31}^d & \sim  2\, y_*^2\, \frac{m_b}{v}\, \left(  \frac{m_d}{m_b \lambda_C^3} \right)
 & \Longrightarrow  \quad
 \Lambda  \gtrsim 570\, \text{GeV}\times y_* \times \sqrt{\frac{200\,\text{GeV}}{m_h}}\, .
\hspace*{2cm} 
\end{align}
Similarly, for $B_s\bar B_s$ mixing we find
\begin{equation}
\begin{split}
\hat y_{23}^d \sim & 2 y_*^2 \left( \frac{m_b}{v}\, \frac{(\lambda_L)_2(\lambda_L)_3}{16\pi^2}  
 + \frac{m_s}{v}\, \frac{(\lambda_R)_2(\lambda_R)_3}{16\pi^2} \right)\\[0.4cm]
\hat y_{32}^d \sim & 2 y_*^2 \left( \frac{m_s}{v}\, \frac{(\lambda_L)_2(\lambda_L)_3}{16\pi^2}
 + \frac{m_b}{v}\, \frac{(\lambda_R)_2(\lambda_R)_3}{16\pi^2} \right)\, ,
\end{split}
\end{equation} 
and the strongest constraint for $b_L$ and $b_R$ fully composite respectively comes
from $C_2\propto (\hat y_{23}^d)^2$ and $\tilde C_2\propto (\hat y_{32}^d)^2$:
\begin{align}
\hspace*{2cm} 
\hat y_{23}^d & \sim  2\, y_*^2\, \frac{m_b}{v}\, \lambda_C^2 
 & \Longrightarrow  \quad
 \Lambda  \gtrsim 135 \, \text{GeV}\times y_* \times \sqrt{\frac{200\,\text{GeV}}{m_h}}
\phantom{\, .}
\hspace*{2cm} 
\\[0.3cm]
 \hat y_{32}^d & \sim  2\, y_*^2\, \frac{m_b}{v}\, \left( \frac{m_s}{m_b \lambda_C^2} \right)
 & \Longrightarrow  \quad
 \Lambda  \gtrsim 420\, \text{GeV}\times y_* \times \sqrt{\frac{200\,\text{GeV}}{m_h}}\, .
\hspace*{2cm} 
\end{align}

\section{Discussion and conclusions}

Our analysis has shown that 
$\Delta F=2$ neutral currents generated by the 
tree-level exchange of a composite Higgs lead to rather strong constraints on the scale of New Physics
if the Higgs is light and strongly coupled. 
We have focussed on scenarios where the 
SM fermions couple linearly to operators of the new strong sector 
that gives the Higgs as a bound state.
We have further assumed that the hierarchy in the SM Yukawa couplings entirely originates
from the RG running of such couplings, while the strong sector is flavor anarchic.

In the case of CP violation in $K\bar K$ mixing the bounds that we have derived 
are comparable to those from the exchange of heavy vectors,
despite the fact that the Higgs exchange requires flavor violation on both vertices
and thus naively counts as a dimension-8 effect.  
We showed that the lightness of the Higgs and its strong coupling  to the EWSB dynamics, $y_*$,
compensate for the naive suppression.
Moreover, while the Higgs exchange is enhanced for larger values of $y_*$, the vector exchange
is suppressed, and in that sense the two constraints are complementary.
Milder bounds follow instead from $\Delta B=2$ transitions.

The above picture is however substantially modified 
in the special and well motivated case of
a composite pseudo-Goldstone  Higgs. In the simplest situations the shift
symmetry acting on the Higgs forces a flavor alignment between the SM Yukawa term 
and the higher-order non-derivative operators with larger powers of the Higgs field. 
Flavor violation then occurs only through the derivative operators, implying
an additional suppression of the flavor-violating Higgs vertex by 
the degree of compositeness of the SM fermions involved.
As a result, the constraints from $\epsilon_K$ are negligible and 
the most stringent bounds come in this case from $\Delta B=2$ transitions.
Moreover, the latter bounds can become as strong
as the corresponding ones for a non-PGB Higgs only 
if $b_L$ or $b_R$ is fully composite.
In that limit however, the constraints from the heavy vector 
exchange are quite stringent and dominate~\cite{Giudice:2007fh, Blum:2009sk}. 
Hence, we conclude that PGB Higgs models are only very mildly constrained by
the Higgs contribution to $\Delta F=2$ processes.

It is worth stressing that our results rely on 
assuming that the strong sector is flavor anarchic and that the 
hierarchical structure of the Yukawa couplings has its origin in the running
of the  linear couplings of the SM fermions.
If any of these assumptions is relaxed, then the estimate of the higher-order operators must be reconsidered.
As an interesting example, consider the case in which the strong sector has an approximate global
$SU(3)^5$ flavor symmetry, broken only by quasi marginal operators $O_u$, $O_d$, $O_e$ with the  quantum
numbers of the SM Yukawa couplings~\cite{Rattazzi:2000hs,Csaki:2009wc}.
In the scenario of ref.~\cite{Rattazzi:2000hs}, all the SM fermions are fully composite, 
and the coefficients of the marginal operators are small and reproduce the hierarchy of the SM Yukawa couplings.
At low energy the theory satisfies the criterion of Minimal Flavor Violation, forcing 
$\tilde y^d \propto y^d+ (y^u y^{u\dagger}) y^d + (y^d y^{d\dagger}) y^d + \dots$, 
$\kappa^q \sim \kappa^{q\prime}\propto 1+ y^u y^{u\dagger} +  y^d y^{d\dagger} + \dots$, 
$\kappa^d \sim 1 + y^{d\dagger}y^d + \dots$,
where numerical coefficients multiplying all the terms have been understood and the dots stand
for terms with more Yukawa insertions.
This implies that the flavor-violating effects from $O_y$ and from the derivative operators of eq.(\ref{eq:otherO}) 
are of  the same order and small.
In the models of ref.~\cite{Csaki:2009wc} instead, the SM fermions are partially composite and the coefficients
of the operators $O_u$, $O_d$, $O_e$ are assumed to be sizable and essentially anarchic.~\footnote{As proposed 
by ref.\cite{Csaki:2009wc}, the approximate flavor symmetry of the strong sector can be smaller than $SU(3)^5$ 
and not all the three marginal operators might be actually needed. This does not change our conclusions however.}
The flavor violation induced by $O_u$, $O_d$, $O_e$ 
feeds into the fermionic sector by splitting the 
anomalous dimensions of the 
fermionic operators,   and it is amplified 
by their  RG evolution down to low energy leading to hierarchical SM Yukawa matrices.
For these models our estimate of the higher-order operators 
goes through essentially unchanged,~\footnote{In fact,  there will be an extra
numerical suppression of Higgs-mediated FCNCs
due to the fact that the relative misalignment between the operator $O_y$
and the down Yukawa matrix arises only at higher order in 
the number of flavor spurions.}
both for a PGB and a generic Higgs. Hence, despite the constrained pattern
of flavor violation in terms of spurions with the quantum numbers of the SM Yukawa couplings,
the low-energy theory is not Minimally Flavor Violating, and the bounds are strong.~\footnote{
Notice on the other hand that in this case, similarly to the MFV theory of ref.~\cite{Rattazzi:2000hs}, flavor
violation in the down sector requires the interplay of both  spurions acting in the up and down sectors.
This is to be contrasted with the anarchic scenario considered in the present paper, where 
flavor violation can arise even from the down sector in isolation.}
This shows how the initial assumptions on the structure of the theory are crucial
for determining the strength of the Higgs-mediated FCNCs.

The effect of the Higgs exchange in $\Delta F =1$ transitions is by far negligible compared to 
that of the $Z$ exchange, due to the Yukawa coupling suppression at the flavor-preserving 
vertex.  Remarkable exceptions to this rule are decay processes in which the Higgs is in the initial
or final state. If the Higgs is light, a quite promising  decay mode is $t\to Hc$,
as first pointed out in ref.~\cite{Agashe:2006wa}. 
Starting from the analog of eq.(\ref{eq:yhatest})
applied to the up-quark sector, a simple estimate
shows that even in the case of a PGB Higgs the $tch$ vertex can be
sizable as long as $t_R$ is maximally composite. In such limit the largest contribution
comes from $\hat y^u_{32}$, 
\begin{equation}
\left(\frac{v}{\Lambda}\right)^2\, \bar t \left[ \hat y^{u\, *}_{23}\, P_L+ 
 \hat y^u_{32}\, P_R \right] c + h.c. \, ,
\qquad y^u_{23} \ll y^u_{32} \sim 2\, y_*^2\, \frac{m_t}{v} \left( \frac{m_c}{m_t\, \lambda_C^2} \right)\, ,
\end{equation}
so that the charm quark is mainly right-handed.
For $\Lambda/y_* = 1\,$TeV we estimate $BR(t\to hc) \sim 1\times 10^{-4}$, which should
be within the reach of the LHC~\cite{AguilarSaavedra:2000aj}.
Interestingly, the above estimate is similar to that for a non-PGB Higgs,
except in that case, due to the contribution of the
non-derivative operator $O_y$, both $\hat y^u_{23}$ and $\hat y^u_{32}$  are comparable in size.

If the Higgs is heavier, the same $tch$ vertex implies a flavor-violating Higgs decay $h\to tc$.
If all the remaining decay widths are as in the SM, the above estimate predicts 
$BR(h\to tc) \sim 5\times 10^{-3}$, 
but larger values can be obtained if the rate to gauge bosons
turns out to be suppressed due to modified couplings of the composite Higgs.
See reference \cite{Azatov} for prospects of observing such a signal at the LHC.

Similar or even larger rates for $t\to hc$ and $h\to ct$ are also predicted 
in the different scenarios of refs.~\cite{Babu:1999me,Giudice:2008uua},
where Yukawa couplings of the light fermions involve higher powers of the Higgs field.
In that case however, only very small shifts are expected
in the flavor-preserving $t\bar th$ coupling.
On the contrary, shifts as large as $v^2/f^2\sim 10-20\%$ in the $t\bar th$ coupling are 
a generic prediction of composite Higgs theories, both for the PGB and the non-PGB 
case, independently on whether the top quark is fully composite or not.

\section*{Acknowledgments}

We would like to thank J.A.~Aguilar-Saavedra, A.~Azatov, G.~Giudice, T.~Okui, L.~Silvestrini, M.~Toharia,
L.~Zhu and especially R.~Rattazzi and R.~Sundrum, for valuable discussions.
K.A. was supported in part by NSF grant No. PHY-0652363.

\section*{Appendix}

We collect here the formulas relative to the $B_d\bar B_d$ and $B_s\bar B_s$ mixing.
Integrating out the Higgs boson generates the $\Delta B=2$, $\Delta S=0$
low-energy effective Lagrangian
\begin{gather}
{\cal L}_{ \Delta S = 0 }^{\Delta B=2} = {\cal O}_2 C_2 + \tilde{ {\cal O} }_2 \tilde{C}_2 + {\cal O}_4 C_4 + h.c. \\[0.5cm]
%%%
{\cal O}_4  \equiv \left( \overline{ b^{\alpha}_L } d_{ R \, \alpha } \right)   
  \left( \overline{ b^{ \alpha }_R } d_{ L \, \alpha } \right) , \qquad
{\cal O}_2 \equiv \left( \overline{ b^{ \alpha }_R } d_{ L \,\alpha } \right)^2   ,\qquad
\tilde{ {\cal O} }_2 \equiv \left( \overline{ b^{ \alpha }_L } d_{ R \, \alpha } \right)^2 \, ,
\end{gather}
with
\begin{equation}
\left( C_4, \; C_2, \; \tilde{C}_2 \right)  =  
\frac{1}{m_h^2} \left( \frac{ v^2 }{ \Lambda^2 } \right)^2 
\left( \hat{y}^d_{ 13 } \hat{y}^{ d \; \ast }_{ 31 }, 
\; \frac{1}{2}\, (\hat{y}^{d\, *}_{ 13 })^2 , \;  \frac{1}{2}\, (\hat{y}^{d}_{ 3 1 })^2 \right)\, .
\end{equation}
The corresponding bound on the Wilson coefficients at the scale $m_h=200 \,\text{GeV}$
which follows from ref.~\cite{Bona:2007vi} is
\begin{equation}
\text{Im} \left( C_4, \; C_2, \; \tilde{C}_2
\right) \, (\mu = m_h = 200\,\text{GeV})  \lesssim 
\frac{1}{ \Big\{ \left( 1.44, \; 0.94, \; 0.94 \right)
\times 10^3 \; \hbox{TeV} \Big\} ^2 }\, .
\end{equation}

The analogous formulas in the case of the $\Delta B=2$, $\Delta S=2$ effective Lagrangian
read
\begin{gather}
{\cal L}_{ \Delta S = 2 }^{\Delta B=2} = {\cal O}_2 C_2 + \tilde{ {\cal O} }_2 \tilde{C}_2 + {\cal O}_4 C_4 + h.c. \\[0.5cm]
%%%
{\cal O}_4  \equiv \left( \overline{ b^{\alpha}_L } s_{ R \, \alpha } \right)   
  \left( \overline{ b^{ \alpha }_R } s_{ L \, \alpha } \right) , \qquad
{\cal O}_2 \equiv \left( \overline{ b^{ \alpha }_R } s_{ L \,\alpha } \right)^2   ,\qquad
\tilde{ {\cal O} }_2 \equiv \left( \overline{ b^{ \alpha }_L } s_{ R \, \alpha } \right)^2 \, , \\[1.0cm]
%%%
\left( C_4, \; C_2, \; \tilde{C}_2 \right)  =  
\frac{1}{m_h^2} \left( \frac{ v^2 }{ \Lambda^2 } \right)^2 
\left( \hat{y}^d_{ 23 } \hat{y}^{ d \; \ast }_{ 32 }, 
\; \frac{1}{2}\, (\hat{y}^{d\, *}_{ 23 })^2 , \;  \frac{1}{2}\, (\hat{y}^{d}_{ 3 2 })^2 \right)\, ,  \\[1.0cm]
%%%
\text{Im} \left( C_4, \; C_2, \; \tilde{C}_2
\right) \, (\mu = m_h = 200\,\text{GeV})  \lesssim 
\frac{1}{ \Big\{ \left( 176, \; 107, \; 107 \right) \hbox{TeV} \Big\} ^2 }\, .
\end{gather}
%

%%%%%%%%%%
%%%%%%%%%%    References
%%%%%%%%%%

\end{document}